\documentclass[aps,prl,twocolumn,amsmath,amssymb,floatfix,nofootinbib,superscriptaddress]{revtex4-1}

\newcommand{\nhat}{\hat{ \mathbf{n}}}

\usepackage{graphicx}
\usepackage{dcolumn}
\usepackage{bm}
\usepackage[usenames, dvipsnames]{color}
\usepackage[normalem]{ulem}

\graphicspath{ {figures/} }

\newcommand{\Bobs}{B^\mathrm{obs}}
\newcommand{\xobs}{x^\mathrm{obs}}
\newcommand{\xtr}{x_\mathrm{tr}}

\newcommand{\be}{\begin{eqnarray}}

\newcommand{\ee}{\end{eqnarray}}

\newcommand{\shrinkfac}{1}

\begin{document}


\title{Reconstructing CMB fluctuations and the mean reionization optical depth }

\newcommand{\cita}{CITA, University of Toronto, 60 St.~George Street, Toronto, ON M5S 3H8, Canada}

\author{P. Daniel Meerburg}
\affiliation{\cita}

\author{Joel Meyers}
\affiliation{\cita}

\author{Kendrick M. Smith}
\affiliation{Perimeter Institute for Theoretical Physics, Waterloo, ON N2L 2Y5, Canada}

\author{Alexander van Engelen}
\affiliation{\cita}



\date{\today}

\begin{abstract}

The Thomson optical depth from reionization is a limiting factor in measuring the amplitude of primordial fluctuations, and hence in measuring physics that affects the low-redshift amplitude, such as the neutrino masses.   
Current constraints on the optical depth, based on directly measuring large-scale cosmic microwave background (CMB) polarization, are challenging due to foregrounds and systematic effects.  
Here, we consider an indirect measurement of large-scale polarization, using observed maps of small-scale polarization together with maps of fields that distort the CMB, such as CMB lensing and patchy reionization.
We find that very futuristic CMB surveys will be able to reconstruct large-scale polarization, and thus the mean optical depth, using only measurements on small scales.  
\end{abstract}


\maketitle

\paragraph{\textbf{Introduction.}}
\vspace{-1.2em}
In the coming decades, multiple planned or proposed experiments aim to measure the polarization of the cosmic microwave background (CMB) sky~
\cite{Henderson:2015nzj,Benson:2014qhw,Suzuki:2015zzg,CMBS4,CLASSexp,Matsumura:2013aja,2015hsa8.conf..334R,Kogut:2011xw} and to map the large scale structure of the universe \cite{Aghamousa:2016zmz, Abell:2009aa, Amendola:2012ys}. Among the primary scientific goals of these experiments is a measurement of the sum of neutrino masses, obtained from a comparison of the amplitude of density fluctuations today with the amplitude at primordial times. The use of the measured CMB anisotropies for the primordial amplitude is subject to a degeneracy with the mean optical depth due to reionization, $\bar{\tau}$ \cite{Zaldarriaga:1997ch,Allison:2015qca}.\footnote{We use $\bar \tau$ for the mean optical depth, using $\tau$ to represent the field in general which can have spatial fluctuations.} 
Given the planned configurations of upcoming CMB experiments and large scale structure surveys, one needs an error on the optical depth of about $\sigma({\bar{\tau}}) \lesssim 0.005$ in order to achieve a $3\sigma$ detection of the sum of neutrino masses in the minimal scenario, i.e., $\sigma\left({\sum m_{\nu}}\right) \leq 0.02$ eV~\cite{CMBS4}.  

The degeneracy between the primordial amplitude and the optical depth can be broken with a measurement of large-scale $E$-mode CMB polarization, which is generated during reionization and as such is directly proportional to the optical depth \cite{Zaldarriaga:1996ke}.  
Determining the optical depth from direct measurements of $E$-mode polarization at large angular scales has proven to be challenging, requiring careful control of instrumental systematics as well as multi-frequency observations to separate Galactic foregrounds.  Before the mission, it was forecasted that measurements from the \textit{Planck} satellite could achieve $\sigma({\bar{\tau}}) =  0.006$, however, after a careful analysis accounting for systematic errors and polarized foregrounds, the most recent constraint has an uncertainty which is larger by about a factor of 2, $\sigma({\bar{\tau}}) = 0.012$ \cite{PlanckTau2016}. Improving measurements of CMB polarization on large angular scales with ground-based experiments is a  challenging proposition, though it is being pursued~\cite{CLASSexp}. A CMB satellite mission seems the most promising way to improve measurements of polarization on large angular scales, but any such mission is likely to be at least a decade away.

It would be extremely valuable if there were an alternative method to obtain a measurement of $\bar{\tau}$ which does not rely on direct measurement of the CMB polarization signal on large angular scales. For instance, it has recently been proposed that measurements of fluctuations in the the 21-cm signal during reionization could be converted, using reionization modelling, into an effective constraint on the mean optical depth  of  $\sigma({\bar{\tau}}) \sim 0.0008$ \cite{21cmTau}. While promising, this constraint would be model dependent.

Here we discuss an alternative method, using measurements of CMB polarization on small scales to reconstruct large scale $E$ modes. On small scales, CMB $B$ modes are primarily sourced by the conversion of $E$ modes into $B$ modes due to two effects: weak gravitational lensing by large scale structure and patchy screening by an inhomogeneous optical depth. Recent CMB surveys have provided maps of CMB $B$ modes induced by lensing \cite{Ade:2014afa,Keisler:2015hfa,Ade:2015tva,Ade:2015zua},  which have been used to obtain lensing maps \cite{Hanson:2013hsb,vanEngelen:2014zlh,Story:2014hni,Ade:2015zua}.  Upcoming surveys will provide maps of both these fields over large fractions of the sky that will be sample variance-limited to small angular scales \cite{Henderson:2015nzj,Benson:2014qhw,Suzuki:2015zzg,CMBS4}.
Given maps of $B$ modes and either lensing or inhomogeneous  optical depth, we show that one can recover the $E$ modes from which the $B$ modes are sourced; this can include $E$ modes on larger angular scales than are directly probed by the experiment. 

The physical processes of how gravitational lensing and patchy screening convert $E$ into $B$ are completely understood, and with a sufficiently low noise experiment our proposal provides a robust method for measuring $E$-mode polarization on large angular scales.\footnote{The power spectrum of inhomogeneous optical depth has not yet been measured \cite{Dvorkin:2008tf,veraInPrep}, leading to some model dependence in our forecasts.  However, once such a measurement is achieved our method will not be subject to any modeling uncertainty.}

\newpage

\paragraph{\textbf{$E$-mode reconstruction.}}
Maps of CMB polarization contain only $E$ modes at linear order in cosmological perturbation theory for a universe with only density perturbations.  However, at second order several distortion fields are known to generate $B$ modes from $E$ modes. First, gravitational lensing by matter between us and the last scattering surface deflects the polarization field in the direction $\nhat$ according to $\widetilde{P_\pm}(\nhat) = P_\pm(\nhat + \nabla \phi(\nhat))$. Second, patchy screening by a variation in the Thomson optical depth modulates the polarization field according to $e^{-\bar \tau}\widetilde{P_\pm}(\nhat) = e^{-\tau(\nhat)}P_\pm(\nhat)$. Here the polarization field is given in terms of the Stokes parameters by $P_\pm(\nhat) = Q(\nhat) \pm iU(\nhat)$, $\phi$ is the lensing potential, and $\tau$ is the spatially-dependent optical depth arising from variations of the ionization fraction or the baryon density.  

Lens-induced $B$ modes were first detected in Ref.~\cite{Hanson:2013hsb}, by cross-correlating a noisy map of $B$ modes with a template $B$ map obtained from $E$--$\phi$ correlation, where the $\phi$ map was obtained by proxy using a large-scale-structure tracer.  Although cast  as a measure of the $B$-mode power spectrum,  this measurement (and the subsequent measurement from \textit{Planck}~\cite{Ade:2015zua}) is effectively a detection of the $\langle E B \phi \rangle$ bispectrum generated by lensing.  Similarly, Refs.~\cite{Hanson:2013hsb,Ade:2013hjl,vanEngelen:2014zlh}  used a $\phi$  map obtained with a $EB$ quadratic estimator in correlation with a $\phi$ tracer to provide effectively an estimate of the  $\phi$ power spectrum (up to a scaling factor relating the $\phi$ maps to the  tracer); this is again a measure of the $\langle E B \phi \rangle$ bispectrum.  We now derive a method which makes use of the same bispectrum to obtain a reconstruction of the $E$ modes given maps of $B$ and $\phi$ (or $\tau$).

A general distortion field $x(\nhat)$, where $x \in \{\phi,\tau\},$\footnote{Other effects such as anisotropic cosmic birefringence~\cite{birefringence2012} or compensated isocurvature modes~\cite{Holder:2009gd,Grin:2011tf} would also convert $E$ into $B$ but require non-standard physics.} will yield $B$ modes of the form 
\be 
	\widetilde{B}_{\ell_1 m_1}^* = \sum_{\ell_2 m_2 \ell_3 m_3} \Gamma^{(x)\ell_1 \ell_2 \ell_3}_{m_1 m_2 m_3} x_{\ell_2 m_2} E_{\ell_3 m_3} \, ,
\ee
where we have expanded to first order in $x$, and
\be 
    \Gamma^{(x)\ell_1 \ell_2 \ell_3}_{m_1 m_2 m_3} &=& (-i) o_{\ell_1 \ell_2 \ell_3} \sqrt{\frac{(2\ell_1 + 1)(2\ell_2+1)(2\ell_3+1)}{4\pi}} \nonumber \\ 
    && \times   J^{(x)}_{\ell_1 \ell_2 \ell_3} \left( \begin{matrix} \ell_1 & \ell_2 & \ell_3 \\ m_1 & m_2 & m_3 \end{matrix} \right) \,
\ee
where $o_{\ell_1 \ell_2 \ell_3}$ is equal to unity when the sum $\ell_1 + \ell_2 + \ell_3$ is odd and zero when the sum is even, with
\be
	J^{(\phi)}_{\ell_1 \ell_2 \ell_3} &=&   \frac{-\ell_1(\ell_1+1)+\ell_2(\ell_2+1) + \ell_3(\ell_3+1)}{2} \nonumber \\ 
	&&\times \left( \begin{matrix} \ell_1 & \ell_2 & \ell_3 \\ -2 & 0 & 2 \end{matrix} \right), \\
	J^{(\tau)}_{\ell_1 \ell_2 \ell_3} &=& e^{-\bar{\tau}} \left( \begin{matrix} \ell_1 & \ell_2 & \ell_3 \\ -2 & 0 & 2 \end{matrix} \right) \, .
\ee

Suppose we have observations of $\widetilde B$ and  $x$, denoted $\Bobs$ and $\xobs$, with power spectra $C_\ell^{BB}+N_\ell^{BB}$ and $C_\ell^{\xtr\xtr}+N_\ell^{\xtr\xtr}$ respectively. In general, $\xtr$ will not be perfectly correlated with $x$. For instance, when $\xobs$ is obtained from the CMB itself using standard reconstruction techniques (e.g., Ref.~\cite{Hu:2001kj} for $x=\phi$ and Ref.~\cite{Dvorkin:2008tf} for $x=\tau$), one makes use of the CMB anisotropies at $\ell > 20$.  Since these modes are sourced at recombination, $z=1100$, and not at reionization around $z\sim 10 $, where the low-$\ell$ $E$ modes are sourced, there is imperfect redshift overlap between the observed distortion field $\xobs$ and the distortion field $x$ responsible for generating $B$ modes from the low-$\ell$ $E$ modes we aim to reconstruct. Alternatively, we may use an external tracer of the distortion field obtained from surveys which may have imperfect redshift overlap with the field of interest \cite{Smith:2010gu}.  We will account for these possibilities by introducing the cross-correlation coefficient~\cite{Simard:2014aqa,Sherwin:2015baa}
\be
	\rho_\ell = \frac{C_\ell^{\xtr x}}{\left[ C_\ell^{\xtr \xtr}C_\ell^{xx} \right]^{1/2} } \,, 
\ee
which will generically differ from unity. 

The formalism for reconstructing $E$ modes follows that developed for lens reconstruction \cite{Hu:2001kj}: we take appropriately weighted combinations of $\Bobs$ and $\xobs$,
\be 
	\widehat{E}_{\ell m} = \sum_{\ell_1 m_1 \ell_2 m_2} W^{(x) \ell_1 \ell_2 \ell}_{m_1 m_2 m} B^{\mathrm{obs}*}_{\ell_1 m_1} x^{\mathrm{obs}*}_{\ell_2 m_2} \, .
\ee
We  choose the weights $W^{(x)}$ to minimize the variance of the reconstruction, subject to the constraint that our estimate be unbiased, i.e., $\left\langle\widehat{E}_{\ell m} \right \rangle = E_{\ell m}$,
\be 
	\sum_{\ell_1 m_1 \ell_2 m_2} W^{(x) \ell_1 \ell_2 \ell}_{m_1 m_2 m} \Gamma^{(x) \ell_1 \ell_2 \ell}_{m_1 m_2 m} C_\ell^{\xtr x} = 1 \, .
\ee
Minimizing the variance gives the optimal weights
\be 
	W^{(x) \ell_1 \ell_2 \ell }_{m_1 m_2 m} &=& N_\ell^{\widehat{E} \widehat{E} (x)}\left(\frac{1}{C_{\ell_1}^{BB,\mathrm{res}} + N_{\ell_1}^{BB}} \right) \nonumber \\ 
    &&\times \left(\frac{C_{\ell_2}^{\xtr x} } {C_{\ell_2}^{\xtr \xtr } + N_{\ell_2}^{\xtr \xtr} }\right) \Gamma^{(x) \ell_1 \ell_2 \ell *}_{m_1 m_2 m} \, 
\ee
and the noise on the reconstructed $E$ modes
\be
	&&N_{\ell}^{\widehat{E} \widehat{E} (x)} =  \left[ \sum_{\ell_1 \ell_2} o_{\ell_1 \ell_2 \ell} \frac{(2\ell_1+1)(2\ell_2+1)}{4\pi} \left(J^{(x)}_{\ell_1 \ell_2 \ell}\right)^2 \right. \nonumber \\
	&& \quad \times  \left. \left(\frac{1}{C_{\ell_1}^{BB,\mathrm{res}} + N_{\ell_1}^{BB}}\right) \rho_{\ell_2}^2 \left( \frac{C_{\ell_2}^{\xtr \xtr} C_{\ell_2}^{x x} }{C_{\ell_2}^{\xtr \xtr}+N_{\ell_2}^{\xtr \xtr}}\right)\right]^{-1}. \nonumber \\ 
    \label{eq:reconstructionnoise}
\ee 

\begin{figure}
\includegraphics[width=\shrinkfac\columnwidth]{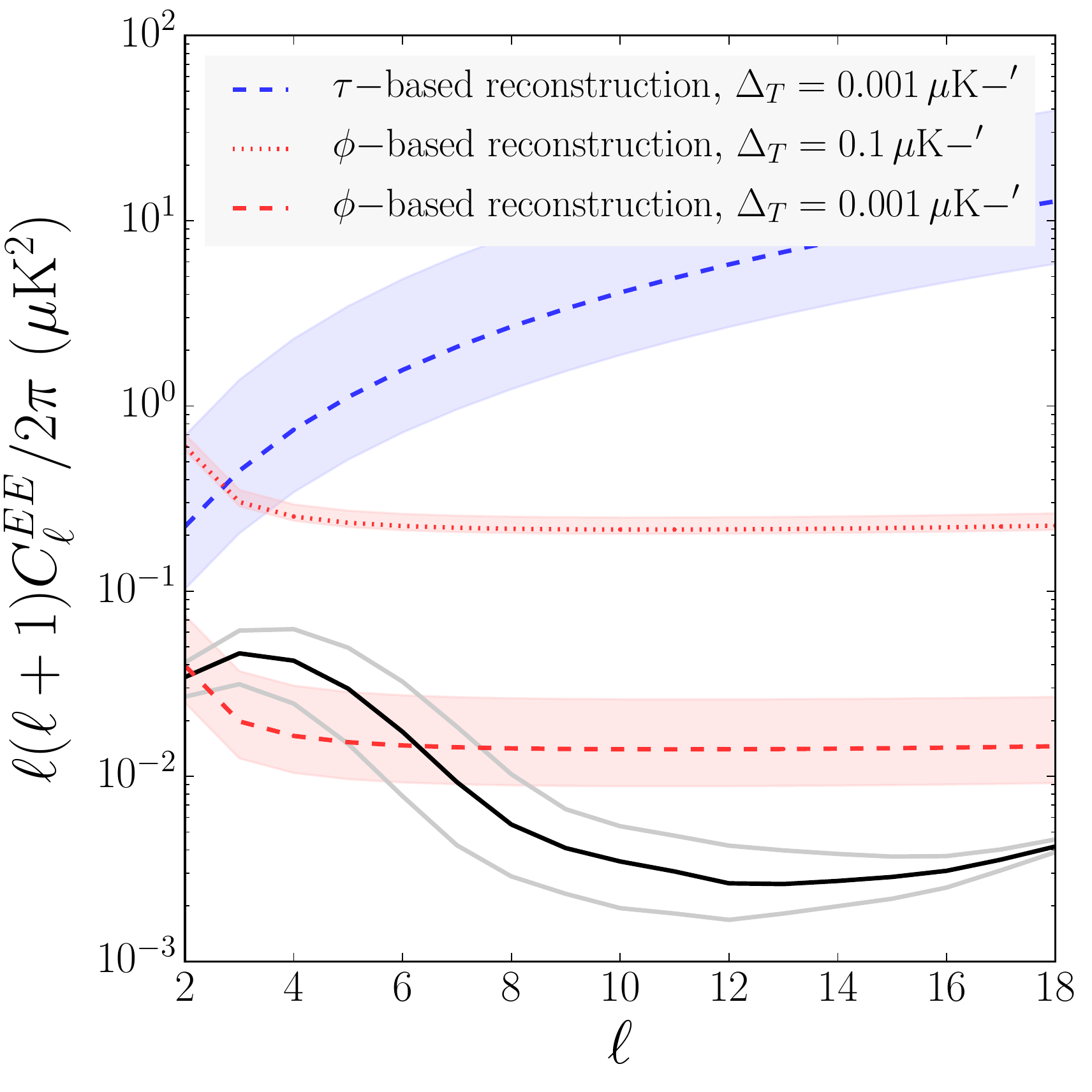}
\caption{Large-scale $E$-mode CMB polarization power spectrum (black) with noise from reconstruction (Eq.~\ref{eq:reconstructionnoise}; blue/red).  We consider reconstructions using either lensing ($\phi$, red) or patchy optical depth ($\tau$, blue) together with their associated small-scale $B$ modes.  Bands reflect the range of reionization models we consider. The black curve is the fiducial model for $\bar \tau = 0.058$ and the grey curves use values offset by current \textit{Planck}\ 1$\sigma$ limits, namely   $\bar \tau = 0.046$ and   $0.070$. }
\label{fig:E_noise}
\end{figure}

The $B$-mode power spectrum has several contributions, including those from primary $B$ modes, from inhomogeneous screening, and from lensing. The reconstruction noise can be reduced by removing the contributions to the $B$-mode power spectrum which come from well-measured $E$ and $x$ modes. To achieve this we iteratively delens and `descreen' the $B$-mode power spectrum assuming no measurement of $E$ modes at $\ell<20$ \cite{Hirata:2003ka, Smith:2010gu}, which serves the dual purpose of reducing the residual $B$-mode power and reducing the noise on the $EB$ reconstruction of the $\phi$ and $\tau$ fields.\footnote{We assume that screened and scattered $B$ modes from patchy optical depth \cite{Hu:1999vq,patchytauCMB} can be reduced in the same way, though a more careful treatment would be required to accurately subtract scattered $B$ modes.} We refer to the residual $B$-mode power after delensing and descreening as $C_\ell^{BB,\mathrm{res}}$. We assume negligible primordial contribution to the $B$-mode power spectrum.

We compute the screened $B$ modes as in Ref.~\cite{patchytauCMB} using a fiducial $C_{\ell}^{\tau \tau}$ with a peak amplitude of roughly $\ell(\ell+1)C_\ell^{\tau\tau}/(2\pi) \sim 3 \times 10^{-6}$ at $\ell \simeq 500$ derived from the simulations of Ref.~\cite{PatchyTauSims}. Given the model uncertainty in reionization~\cite{Alvarez:2010ht,Alvarez:2015xzu}, we also scale this fiducial model by a factor of 3 in both directions. 

\begin{figure}
\includegraphics[width=\shrinkfac\columnwidth]{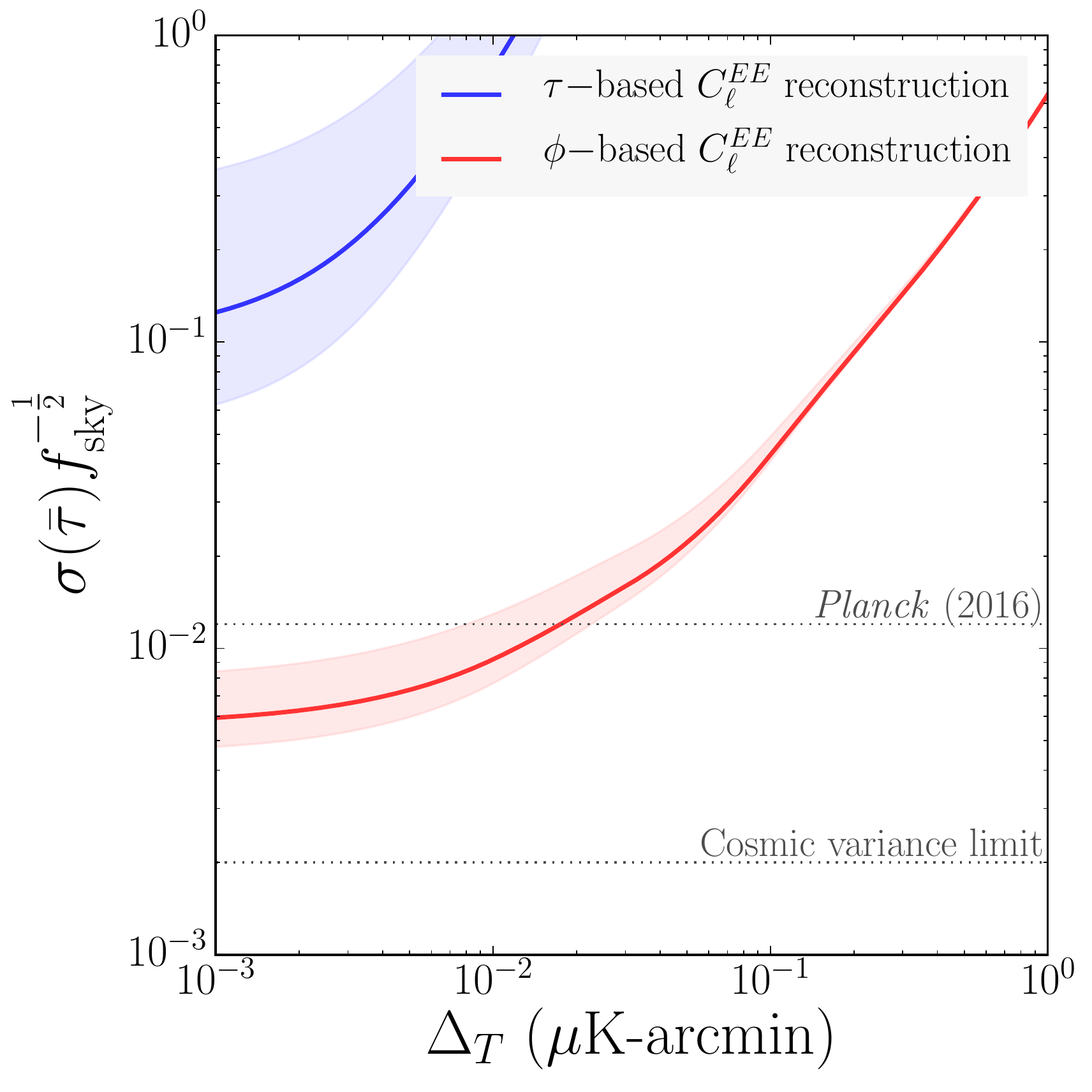}
\caption{Forecasted constraints on the mean optical depth, $\bar \tau$, when using small-scale measurements of $B$ and $x = (\phi, \tau)$ to reconstruct $E$ on large scales $\ell < 20$.   We assume that both the $B$ modes and the reconstructed $x$ modes have been obtained from a CMB survey with the given temperature noise level, $\Delta_T$, at  $\ell > 20$ and a 1 arcmin beam.  As in Fig.~\ref{fig:E_noise}, bands indicate the range of results depending on the reionization model we consider. The dotted lines indicate the current constraint, from Ref.~\cite{PlanckTau2016}, and the CMB cosmic variance limit.  Although the $\tau$-based constraint is not competitive, we note that it could be improved using a large-scale structure tracer as a proxy.}
\label{fig:tau_constraints}
\end{figure}

For our forecasts, we use internal CMB reconstruction of the $\phi$ and $\tau$ fields, assuming a 1 arcmin beam and no measurement of large scales $\ell<20$.  For simplicity, we assume that the reionization $E$ modes are sourced on a single screen at $z \simeq 10$.  The lensing potential field for sources at $z \simeq 10$ is highly correlated with that for sources at $z = 1100$, having a cross-correlation coefficient $\rho_\ell \gtrsim 0.9$ on all scales.  The situation for patchy reionization is more complicated, since the source redshifts inevitably overlap with the screening redshifts. We conservatively consider only  fluctuations in the ionized gas after reionization, at $z<6$, to be responsible for screening the reionization  $E$ modes \cite{Battaglia:2016xbi,veraInPrep}. We find that this low-redshift component has a cross-correlation coefficient with the total inhomogeneous optical depth field, including both reionization and low-redshift contributions, ranging between $\rho_\ell = 0.1$--$0.4$ depending upon the reionization model~\cite{veraInPrep}.

In Fig.~\ref{fig:E_noise}, we show the $E$-mode reconstruction noise  compared to the $C_\ell^{EE}$ power spectrum for several currently allowed values of $\bar{\tau}$. We find that the $B\phi$ reconstruction noise is comparable to the signal only for  very low noise experiments. The $B\tau$ reconstruction noise is larger for all noise levels. We considered screening of reionization $E$ modes only by low-redshift optical depth sources; lower noise could be achieved by treating screening as a continuous process throughout reionization~\cite{Dvorkin:2008tf}, or by using a tracer for the $\tau$ field. Models with a lower amplitude for the reionization component of $C_{\ell}^{\tau \tau}$ also result in lower reconstruction noise.

\paragraph{\textbf{Constraints on the mean Thomson optical depth.} }
For each noise level and type of reconstruction we determine the error on the sky-averaged optical depth $\bar{\tau}$ using~\cite{Knox:1995dq,Zaldarriaga:1997ch}
\begin{equation}
\sigma(\bar{\tau}) = \left [ \sum_{\ell=2}^{\ell_\mathrm{max}}\left( \frac{\partial C_\ell^{EE} }{\partial \bar{\tau}}\right)^2\frac{1}{\sigma^2 \left(C_\ell^{\widehat{E}\widehat{E}}\right) }\right]^{-1/2}\, ,
\end{equation}
with
\be
\sigma^2 \left(C_\ell^{\widehat{E}\widehat{E}}\right) = \frac{2}{(2\ell + 1)f_\mathrm{sky}} \left(C_\ell^{EE} + N_{\ell}^{\widehat{E} \widehat{E}(x)}\right )^2\, ,
\ee
where $f_{\mathrm{sky}}$ is the surveyed sky fraction.
In Fig.~\ref{fig:tau_constraints} we show associated constraints. 

	Unfortunately, the forecasted error on $\bar \tau$ from reconstruction is lower than the \textit{Planck} result only for noise levels of $\Delta_T \lesssim 2\times 10^{-2} \,\mu$K-arcmin, much lower than the expected CMB-S4 survey noise of $\Delta_T \simeq 1  \, \mu$K-arcmin, preventing us from constraining $\bar{\tau}$ by reconstruction  with any currently planned survey. 
Given the current central value of $\bar{\tau}  = 0.058$ we estimate a noise level $\Delta_T \lesssim 5\times 10^{-2} \,\mu$K-arcmin would be required for a 3$\sigma$ detection of the mean optical depth.

\paragraph{\textbf{Discussion and conclusions.}}
The Thomson optical depth from reionization is a limiting factor in measuring the amplitude of primordial fluctuations, and hence in measuring physics that affect the low-redshift amplitude, such as the  neutrino masses. The most obvious path forward is to build a CMB satellite capable of measuring the large angular polarization pattern on the sky to cosmic variance limits. Here we presented an alternative; late time physics in the form of gravitational lensing and patchy reionization converts large angle $E$ modes to small angle $B$ modes, which can be measured to exquisite precision from the ground. We developed the tools to use these measurements to reconstruct  the original $E$ mode signal, which can then be used to constrain the mean optical depth.

Our forecasts showed that it is possible to reconstruct the large scale $E$ modes using measurements of small scale polarization, though it would require a survey with extremely low noise, $\Delta_T \lesssim 2\times 10^{-2} \, \mu$K-arcmin, in order to improve upon current constraints.  Maps of reconstructed $E$ modes could also be cross-correlated with direct measurements of the low $\ell$ E-mode polarization to mitigate systematic effects and astrophysical foregrounds which may bias constraints from direct measurements of $E$-mode polarization.

The $E$-mode reconstruction noise is limited by residual $B$-mode power, the noise of reconstructed distortion fields, and imperfect redshift overlap between the reconstructed distortion fields and those acting on the reionization $E$ modes. It may be possible to alleviate these limitations and improve $E$-mode reconstruction by using external tracers of the distortion fields. For example, observations of the cosmic infrared background, galaxy catalogs, or 21-cm surveys may provide useful probes of the large scale structure responsible for lensing the CMB \cite{Song:2002sg,2006ApJ...653..922Z,Bleem:2012gm,Sherwin:2012mr,Ade:2013aro,Pourtsidou21032014,Giannantonio:2015ahz}.  Similarly, a map of large-scale structure could provide a useful proxy for the low redshift contributions to the inhomogeneous ionized gas. 

Our analysis only considered estimators involving secondary $B$-mode polarization. However, on very small scales (i.e. $\ell >3000$), secondary $E$ modes from lensing dominate over the primary CMB contributions, and have power equal to that of $B$ modes. We could derive $E\phi$ and $E\tau$ estimators for large scale $E$ modes, potentially lowering the reconstruction noise by up to a factor of 2 when using delensed small scale $E$ modes \cite{Green:2016cjr}.

We assumed that reionization $E$ modes were sourced at a single redshift and only screened by low-redshift sources of inhomogeneous optical depth.  A more complete calculation would treat the generation of polarization and screening thereof as a continuous process~\cite{Dvorkin:2008tf}, rather than as two discrete events. Our simplified treatment suggests that $E$-mode reconstruction based on patchy screening is not competitive with that of lensing.  However, if low-noise patchy screening maps become available, from an external tracer for example~\cite{Battaglia:2016xbi}, a more careful treatment is warranted.

The reconstruction of CMB modes has other applications: it could be applied to both temperature and polarization on large scales to test for anomalies \cite{Ade:2015hxq,Schwarz:2015cma}. 

In this short paper we proposed a technique for reconstructing large scale CMB fluctuations using only small scale CMB observations. We showed that it is possible with this method to measure the mean Thomson optical depth with a precision sufficient to guarantee a high significance detection of the sum of neutrino masses, though achieving such a measurement would require a CMB survey with noise significantly below any which is currently being planned. Improvements such as the use of external tracers may help to reduce the technical requirements and make the reconstruction method discussed here a viable path toward constraining the mean optical depth.

\vspace{.4cm}
\noindent
\paragraph{\textbf{Acknowledgements.}}
We would like to thank Marcelo Alvarez and Nick Battaglia for discussion and for providing simulated $C_\ell^{\tau\tau} $ power spectra. We also thank Neelima Sehgal for  discussion.\ JM was supported by the Vincent and Beatrice Tremaine Fellowship.KMS was supported by an NSERC Discovery Grant and an Ontario Early Researcher Award.

\newpage

\bibliography{tau,cmbs4}

\end{document}